\documentclass{article}
\usepackage{txfonts}\usepackage{balance}\usepackage{arxiv}\usepackage{natbib}
\usepackage{graphicx}
\usepackage{float}
\usepackage[a4paper]{hyperref}
\begin{document}

\title{Splinter Session ``Solar and Stellar Flares''}
\author{L. Fletcher$^1$, H. Hudson$^{1,2}$, G. Cauzzi$^3$, K. V. Getman$^4$, M. Giampapa$^5$, S. L. Hawley$^6$, P. Heinzel$^7$, C. Johnstone$^8$, A. F.  Kowalski$^6$, R. A. Osten$^9$ and J. Pye$^{10}$}

\affil{$^1$ School of Physics and Astronomy, SUPA, University of Glasgow, Glasgow G12 8QQ, UK (lyndsay.fletcher@glasgow.ac.uk)}
\affil{$^2$ Space Sciences Laboratory, U. C. Berkeley, 7 Gauss Way, Berkeley, CA 94720, USA}
%%GC
\affil{$^3$ INAF - Osservatorio Astrofisico di Arcetri, Largo Enrico Fermi 5, 50125 Firenze, Italy}
\affil{$^4$ Department of Astronomy \& Astrophysics, 525 Davey Laboratory, Pennsylvania State University, University Park, PA 16802, USA}
\affil{$^5$ National Solar Observatory, 950 North Cherry Ave, Tucson, AZ 85719, USA. }
\affil{$^6$ Astronomy Department, University of Washington, Box 351580, Seattle, WA 98195, USA}
\affil{$^7$ Astronomical Institute, Academy of Sciences of the Czech Republic, 25165 Ond\v rejov, Czech Republic}
\affil{$^8$ School of Physics and Astronomy, University of St Andrews, St Andrews, Scotland KY16 9SS}
\affil{$^9$ Space Telescope Science Institute, 3700 San Martin Drive, Baltimore, MD 21218, USA}
\affil{$^{10}$ Department of Physics and Astronomy, University of Leicester, Leicester, LE1 7RH, UK}

%\authorrunning{Fletcher et al.}
%\titlerunning{Solar and Stellar Flares}

\begin{abstract}
This summary reports on papers presented at the \textit{Cool Stars--16} meeting in the splinter session ``Solar and Stellar flares.''
Although many topics were discussed, the main themes were the commonality of interests, and of physics, between the solar and stellar flare communities, and the opportunities for important new observations in the near future.
\end{abstract}

%\maketitle{}

\section{Introduction}
For several years, studies of solar flares and of stellar flares have -- by necessity -- evolved along diverging tracks. To be observable on an essentially unresolved stellar source, stellar flares must be bright, so that it is possible to make extremely high caliber spectroscopic observations, but without much spatial information.
 On the other hand, the solar observations now provide exquisitely detailed spatial information, so that the spatial evolution of a solar flare is typically very well characterized, but the difficulty of predicting its location means that good spectroscopic information, from a slit placed exactly over the flare, is rare -- especially in the crucial ``impulsive phase.'' 
This splinter meeting, which drew together participants from the solar and stellar flare communities (plus others) aimed to summarize the current state of knowledge in the field, and plot a possible way ahead. 
These notes do not cite the literature comprehensively, but instead mainly reflect current research work, and the topics discussed at the Splinter meeting.

\section{An overview of solar flares}

\begin{figure}
\includegraphics[width=\textwidth]{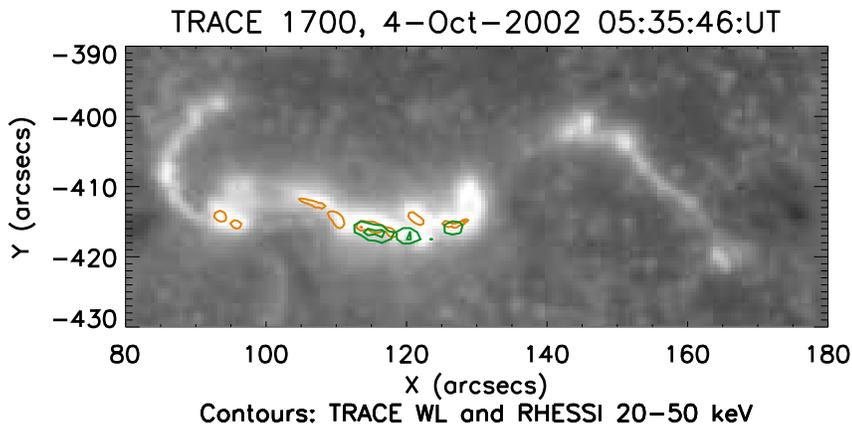}
\caption{High spatial resolution image of a solar flare. The grayscale is the TRACE 1700\AA~channel, orange is the TRACE white light channel and green is RHESSI 25-50~keV footpoints. The field of view spans around 72,000 $\times$ 29,000~km$^2$.}
\label{fig:solarflare}
\end{figure}

Spatially-resolved flare observations in the modern era reveal that the energetically dominant optical and UV emission of a flare, occurring during the few minutes of the flare's impulsive phase, originates in two or more ribbons which spread across the chromosphere. The strongest optical emission is  organized into very compact footpoints,  smaller than a few thousand km on the Sun (see Figure~\ref{fig:solarflare}) which evolve rapidly in position and intensity.
This reflects the progression of the main flare energy release.  Compared to chromospheric and transition-region  ribbons, the optical footpoints have been very little studied in the recent era. White-light (WL) emission had previously been thought to be a Ôbig flareÕ phenomenon, but has now been seen even in small C-class events \citep{2006SoPh..234...79H,2008ApJ...688L.119J}.
WL footpoints have a very precise timing and spatial relationship to flare hard X-rays \citep{2003ApJ...595..483M,2006SoPh..234...79H, 2007ApJ...656.1187F,2010ApJ...715..651W}. Hard X-rays are generated by electrons accelerated during a flare, which are a major energy-carrying component, and the peak of the flare energy spectrum, as far as we can pin it down, is in the optical-UV regime \citep{2010MmSAI..81..637H}. The implication is that energy carried by flare electrons is eventually radiated in the lower-atmospheric optical-UV, even though the energy has been built-up and stored in coronal magnetic structures.
Studies of flare energetics in the two largest flares of the last cycle indicate that the energy contained in solar flare electrons and ions is of the order of $10^{30-32}$ ergs in each component, comparable to the kinetic energy of the coronal mass ejections associated with the events, and with the estimates of the free magnetic energy extrapolated from the photospheric boundary \citep{2004JGRA..10910104E,2005JGRA..11011103E}.
Non-thermal electrons are of core importance in solar flares, and the results from the \textit{Ramaty High Energy Solar Spectroscopic Imager} \citep[RHESSI;][]{2002SoPh..210....3L} have added substantially to our understanding of this.
RHESSI  provides imaging spectroscopy from 3~keV up into the $\gamma$-ray range, with unprecedented dynamic range and spectral resolution.
It thus uniquely complements the more recent fleet of solar spacecraft, none of which provide this key high-energy data; the latest addition, the \textit{Solar Dynamics Observatory}, only reaches the EUV wavelength range but provides remarkably detailed and complete context information.

Interpreted as the chromospheric intersections of evolving magnetic (quasi-) separatrix surfaces,chromospheric flare ribbons allow the progression of field restructuring to be followed \citep{2009ApJ...700..559M}. Chromospheric flares have of course long been beautifully observed in H$\alpha$, but also now in other lines such as the 304~nm filter from the \textit{Solar Dynamics Observatory} (SDO). 
These show abundant fine detail from 3D structures. Flare ribbons give insight into both energetics and magnetic restructuring, in the impulsive and the gradual phase. The fact that there is a small subset of the ribbons illuminated in optical and HXR indicates that there are special locations in the magnetic field where very efficient electron acceleration takes place. It is not clear what the significance of these sites is; they may be associated with singular structures in the field called `magnetic spines' \citep{2009ApJ...693.1628J}. Flare footpoints are also visible emitting in soft X-rays, at temperatures in the range 8-10~MK, with emission measures consistent with chromospheric densities and volumes \citep{2004A&A...415..377M}.

It is not clear whether the optical footpoint emission corresponds to heating of the photosphere, but during the impulsive phase there is most certainly evidence that the effects of a solar flare can reach this layer. The `sunquakes' \citep{1998Natur.393..317K,2005ApJ...630.1168D}, and the 
%% GC 
non-transient
changes in the photospheric line-of-sight magnetic field \citep{1992SoPh..140...85W,1999ApJ...525L..61C,2005ApJ...635..647S}, though not visible in all flares, demonstrate the photosphere is mechanically perturbed, by a shock, by the Lorentz force, or by some other means.

In response to the impulsive-phase heating the chromosphere expands into the corona, seen spectroscopically in the extreme UV \citep{2009ApJ...699..968M}, with a chromospheric downflow that approximately balances momentum \citep{1988ApJ...324..582Z,2006A&A...455.1123T}. 
The upflow speed is observed to vary with the (LTE)  temperature of the emitting plasma, 
with higher temperature plasmas moving faster.  Arcades of X-ray and EUV loops form, straddling the photospheric polarity inversion line between the chromospheric ribbons. 
As the ribbons spread away from this line, the loops become oriented more perpendicularly to it \citep{1992PASJ...44L.123S}.
This suggests that later phase energy release occurs in magnetic field which -- in addition to weakening -- may become less sheared. 
The loops cool from coronal temperatures, eventually becoming visible in H$\alpha$. The indication is that, at least during larger events, the temperature evolution is slower than would be suggested by radiative and conductive cooling \citep{1998A&A...337..911H} indicating the presence of energy input even during the gradual phase.

Spatially-resolved information provides a clear picture of the flare evolution, and indicates also the physical processes at work. It is an embarrassment that there is no optical or UV imaging flare spectroscopy from space, although the IRIS instrument will soon be available. 
There are also precious few ground-based spectra -- extensive measurements have not been made since the 1980s and early 90s.  The review in this splinter by Cauzzi highlighted some past and present work in ground-based spectroscopy of flares. Recently, this has predominantly been high resolution slit spectroscopy on a small number of spectral lines, e.g. the observations of  H$\alpha$, H$\beta$, Ca H and K, Ca II $\lambda$ 8542 with the refurbished instrument at Ondrejov  \citep{2007ASPC..368..559K}, and lower resolution multi-channel spectrograph work, in which images can be made simultaneously at several points across a spectral line, or a number of spectral lines \citep[e.g.][]{2007A&A...461..303R}, including a flare presented which was observed
%% GC add
in August 2010
with the IBIS instrument \citep{2006SoPh..236..415C}. Observations over a broad spectral range have been very few and far between in recent times \citep{1997ApJS..112..221J}. The chromospheric lines are potentially rich in diagnostic information. For example, \cite{1990ApJ...350..463M,1990ApJ...365..391M}  have used Mg~I lines to probe the temperature and density structure of the temperature minimum region. Both the Ca infrared triplet 
at $\lambda\lambda$ 8498, 8542, 8662 and the Balmer lines provide information on electron density \citep[e.g.][]{1976sofl.book.....S,2002SoPh..207..125D}, and potentially also diagnostics of non-thermal electron distributions \citep{1993SoPh..143..259Z,2001MNRAS.326..943D,2002A&A...382..688K}.

\begin{figure}
\begin{center}
\includegraphics[width=0.5\textwidth]{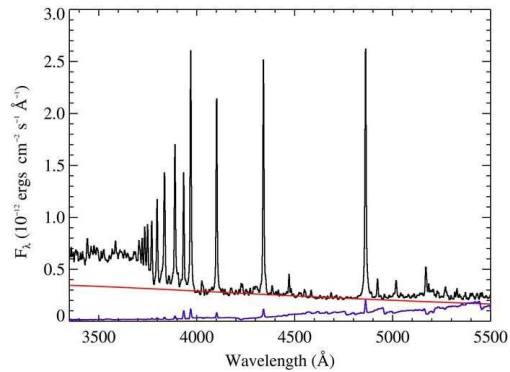}
\caption{High spectral resolution observation of a megaflare on YZ CMi \citep{2010ApJ...714L..98K}: red line, a blackbody fit longward of the Balmer jump; purple line, the preflare stellar spectrum.
}
\label{fig:kowalski}
\end{center}
\end{figure}

HMI on SDO is now providing imaging spectroscopy of the photospheric  Fe I 6173 \AA\ line, and SDO/EVE provides full EUV spectra of the Sun as a star. In the meanwhile, Figure~\ref{fig:kowalski} suggests the power of what may emerge from better spectroscopy: this is a stellar flare spectrum dominated by the Balmer series, but with white-light continuum (and photospheric lines) in the background \citep{2010ApJ...714L..98K}. We do not have modern spectra for solar flares with quality approaching  stellar ones.

\section{An overview of stellar flares}
%similarities
The Sun is a weak flare source. Its flaring energy output is tiny compared to what is seen on other stars \citep{2000ApJ...529.1026S}. 
Even among the `normal' single stars (leaving aside pre-main sequence stars), stellar flares can involve the release of $10^6$ times more energy in total than solar flares, suggesting substantially higher magnetic fields and larger flaring volumes (the available energy is $\int (B^2 - B_{pot}^2)\ dV$, where $B_{pot}$ is the potential field matching the same boundary conditions). 
Flares are observed in all types of stars with convective envelopes, including from M~dwarfs, solar analogs, RS CVn stars, and also giants. 
Flare emission emerges in the optical and near-UV, as well as X-rays and radio but, as with the solar case, a large fraction of the radiated energy for dMe flares emerges in the optical part of the spectrum \citep{1991ApJ...378..725H} during both the impulsive and gradual phases. 
In many cases, the stellar flares observed so far show a great deal of consistency with solar observations, i.e. most of the power is in the continuum, the ratios of radio to UV emission are similar \citep{2004ApJS..153..317O}, the impulsive phase can show similarly abrupt rises to the solar case (minutes) and the decay phases typically last for hours, with evidence for ongoing heating during the decay. 
Hard X-rays, above about 10~keV, have also been observed in stellar flares (though only in a few extremely large ones) and the ratio of hard and soft fluxes follow the same scaling relationship in solar and stellar flares \citep{2007A&A...472..261I}, pointing towards the hard photons having a non-thermal spectrum. 
Solar flare HXR photons account for around $10^{-5}$ of the total flare energy; if the same is true for stellar flares it is no surprise that thus far good stellar flare hard X-ray spectroscopy has not been possible.  The `superflare' on II~Peg (an ``X4.4$\times 10^6$'' event) reported by \cite{2007ApJ...654.1052O} and observed with {\it Swift}/BAT is an exception. While in this event an unambiguously non-thermal spectral shape is not confirmed, it appears highly plausible. 
HXR spectroscopy of smaller events should become possible with the proposed Simbol-X mission. However, observations of gyrosynchrotron emission in stellar flares, which is in fact the principal diagnostic of non-thermal electrons in stellar flares, has long led us to expect that high energy, hard spectrum, non-thermal electrons are present \citep{2002ARA&A..40..217G}; indeed \citep{1996ApJ...471.1002G} found that stellar flares on UV Ceti are radio-overluminous with respect to their X-ray emission in comparison with solar events.

% differences
However, there are observed differences, for example in the shape of light curves.  During the session an example was given by Pye et al. of an XMM Newton flare (on 2MASS 04072181-1210033) with a comparatively rapid decay, and the Chandra Orion Ultradeep Project (COUP) survey of pre main-sequence stars by \cite{2008ApJ...688..418G} reports many examples of slow-rise events with extended flat tops and very long decays, continuing for days.  The soft X-ray rise timescale of a solar flare is characterised by the duration of the impulsive phase (usually a set of discrete energy input events) and the length of the coronal structure to be filled by evaporation -- in other words by the configuration of the magnetic field in which the flare event takes place. These different Chandra characteristics imply a very much larger flaring structure: \cite{2008ApJ...688..418G} suggest they are enhanced analogues to solar long duration events. 

%spatial information
There are also flares showing much more exotic X-ray light curves, with dips and multiple peaks, and it is conceivable that these could be used to deduce spatial information about stellar flares, which is at a premium.  For example, the time variations in about a quarter of the SXR light curves in the COUP survey  \citep{2008ApJ...688..418G} can be explained by the rotational eclipsing of large emitting loops. Work by Johnstone et al. presented in this session identified events in this survey having a significant dip in the lightcurves, constituting about 30\% of all of the flares in the survey. An estimate can be made of the expected number of eclipsing flares by sampling at random the distributions of flare loop lengths and durations (relative to rotation period of the star), estimated from SXR decays by \cite{2008ApJ...688..418G}. This results in a much smaller estimate, i.e. that about 6\% of all flares should show the effect of eclipses in their light curves (the larger, longer-lived ones). The implication is that eclipses are important even in flares of `normal' size and duration'. The effects might be more subtle -- small dips, a reduced peak emission measure compared to an uneclipsed flare, or a shorter apparent duration. How to disentangle the behavior due to eclipsing from the intrinsic flaring behavior is not clear. Some independent estimates of flare X-ray sizes may also be available from the technique of coronal seismology, such as is frequently used now in solar flare studies. For example, variations in the X-ray light curve of a flare loop on AT Mic were interpreted by \cite{2005A&A...436.1041M} as damped oscillations of a standing magneto-acoustic wave, giving constraints on the loop length of $2.5\times 10^{10}$cm and field strength of 100$\pm$50 G. 

Given the spectral flux distribution it is also possible to estimate the size of the excited patch in the flare star chromosphere/photosphere. This has been done under the assumption that the optical flare corresponds to a patch of blackbody emission with an elevated temperature compared to the surrounding photosphere \citep{1992ApJS...78..565H,2003ApJ...597..535H}. Area coverage obtained in this way is typically small -- on the order of 0.01\% to 0.1\% of the star's visible hemisphere (around a factor ten larger than the area coverage of optical patches in solar flares), though larger events are possible (e.g. 3\%  coverage in the Osten et al (2010) `superflare'). The energy input implied by these areas and the luminous energy can be on the order of $10^{11}{\rm erg\ cm^{-2} s^{-2}}$ \citep{2003ApJ...597..535H,2010ApJ...714L..98K,2010ApJ...721..785O} corresponding to an intense solar flare. Evidence was also presented in this session for sub-structure in the spatial distribution of flare optical emission, namely the YZ CMi flare reported by \cite{2010ApJ...714L..98K} analysis of which implies a Balmer continuum-emitting patch which is around 3-15 times larger than the blackbody patch. This may indicate the same behavior as is consistently observed in solar flares, namely relatively extended UV and H$\alpha$ emitting ribbons compared to optical and HXR footpoints.

How invariable is the flare emission spectrum from star to star, considering all wavelengths?
\cite{2005A&A...431..679M} find stellar flare UV and soft X-ray fluxes to correlate, and have comparable energies.
On the other hand the first true bolometric observations of solar flares \citep{2004GeoRL..3110802W} suggest $L_X/L_{bol} \sim 0.01$, which would be quite different.
A partial explanation may be that the solar soft X-ray photometry normally refers to GOES, which does not detect long-wavelength soft X-rays as well as typical stellar instruments.
In contrast to this possible suggestion of a potential unification of flare spectra, \cite{2005ApJ...621..398O} find a surprising variety of flare detectability in disparate wavelength bands.
Even with the same basic physics one would expect systematic differences across stellar variations in scale, abundances, and gravity, and so the global energetics of solar and stellar flares will be an interesting topic for the future.

\section{Surveys and Catalogs}
The Sun and flare stars present us with a wealth of events. On the Sun, we have catalogues of large numbers of flares, including collections of flares from individual active regions. These, and the spatial information available, permit us to generate sets of flares which have been selected for particular characteristics (e.g. flares without an eruption, flares with coronal sources, flares with only two footpoints.... etc). In principle it is possible also to identify sets of flares from an active region with very similar spatial characteristics -- the so-called `homologous flares'. Thus one can try and control the large number of variables which might affect the process of flare evolution, particle acceleration, etc. On the other hand, with the Sun we are restricted to events on a single class of object -- a middle-sized, middle-aged, isolated, main-sequence star. The population of flare stars provides a richer range of flaring environments, as well as clues to how frequent flaring activity is in solar-type, and other types of stars. The first step in such analysis is the survey and cataloging step. The COUP survey mentioned above is one example as is the XMM-Newton Serendipitous Source catalogue presented in this session by Pye. This catalogue, obtained by analysis of the light curves of sources present in the field of XMM-Newton turned up around 130 flaring events, 40 of which came from previously unknown flare sources. 
Preliminary analysis by Pye et al. has shown that the serendipitous sources which flare more than once spend up to 10-20\% of their time in a flaring state, only a factor two or so less than the `known' flare stars among the XMM-Newton sample. 
Furthermore, a few percent of all stars for which an XMM-Newton light curve could be extracted show flaring activity, demonstrating that significant flaring is not so infrequent. As well as the light curves, some spectral information is available (Pye \& Rosen, elsewhere in these Proceedings), represented for this analysis as the ratio of X-ray count rates in two XMM bands (0.2-1, 1-12 keV), giving a clue to the temperature evolution of the flare (assuming a thermal spectrum) -- the kind of analysis which can also be carried out straightforwardly for solar flares based on their GOES X-ray light curves. Large and growing serendipitous catalogs also offer the opportunity to search for flaring activity in solar-type stars.

In the solar-flare arena, work has tended to proceed in two ways: catalog-based statistical analyses of moderate or large populations of flares, and detailed, often multi-wavelength analyses of spatial, spectral and temporal evolution of individual events. It has so far proved difficult to merge the two approaches, detailed studies proving so intricate and throwing up many different facets of flare behavior which would be very time-consuming to analyze in a statistically-significant sample. But the detailed studies provide general information which we believe to be of wide applicability. For example, imaging analysis of the energy input (i.e. the HXR evolution) of a small number of large, long-duration solar flares reveals without a doubt that they are due to a group, possibly a sequence, of events occurring at different locations throughout a spatially extended region, rather than one single loop or site undergoing repeated flaring activity \citep[e.g.][]{2005ApJ...625L.143G}. However, it must be said that the sheer richness of solar data is very tempting; there are relatively few statistical studies of simple properties of large populations compared to studies of detailed and complex evolution of individual flare events. It is clear that for comparisons with stellar populations, more effort has to be expended on the former type of solar flare analysis, informed also by stellar flare work.

\section{Modelling}
Solar and stellar flare modeling have many aspects in common, perhaps because stellar flare theorists and observers have often looked to theories for the better-observed solar flares for initial guidance. 
The beam-driven radiative hydrodynamic approach of \cite{1999ApJ...521..906A} has been applied to solar flares and to dMe star flares as well \citep{2006ApJ...644..484A}, and remains the most sophisticated treatment of the impulsive phase and its consequences. 
Also the analysis of the cooling phase in terms of radiative and conductive losses calls on the same basic physics (and arrives at the same conclusion that there is often evidence for ongoing energisation in the late phase). Where solar modeling is required to be more detailed than stellar flare modeling is in the interpretation of the spatial structure of the flare, and in particular its relationship to the connectivity or topology of the solar magnetic field \citep{1958IAUS....6..123S}. Here a whole area of research has grown up which seeks to describe the three-dimensional magnetic topology of coronal structures rooted in the observed photospheric magnetic field, and to explain the evolution and the energy release in a flare as a consequence of flux transfer from one magnetic domain to another, facilitated by magnetic reconnection \citep[e.g.][]{1987A&A...185..306H,1992SoPh..139..105D,1999SoPh..190....1P}. From such work we have learned the important of magnetic structures such as separators, separatrices and nulls. 
The topology of stellar magnetic fields is also being approached using extrapolations based on Zeeman and Doppler imaging, and particularly well-studied stars reveal topologies significantly more complex than a straightforward dipole  \citep[e.g.][]{2001LNP...573..207D,2002MNRAS.333..339J,2009ApJ...704.1721P,2010MNRAS.407.2269M}. However, it should be cautioned that -- based on the solar experience -- the topology of the magnetic field on the scale of the whole star may have very little to say about the location and evolution of a flare. In the solar case it seems that we will have to understand the configuration and evolution of the non-potential magnetic field on the scale of less than ten thousand kilometers, and on the timescale of the impulsive phase, to be able to map out the locations of energy storage, the likely sites of a flare initiation or trigger, and the subsequent magnetic field restructuring.

In both stellar and solar flare cases, spectroscopic modeling tends to proceed along the same lines, again relying extensively on the radiation hydrodynamic simulations mentioned above. These complex simulations, however, are limited in the number of species and ionization stages that can be included, so other approaches include computing the state of the flare atmosphere under particular assumptions about energy input and redistribution, then using this as input for other spectral synthesis codes - such as the CHIANTI software \citep{1997A&AS..125..149D} for the optically-thin regime (which at present makes the assumption of LTE), or the multi-level non-LTE radiative transfer approach \citep{1991A&A...245..171R,1992A&A...262..209R,1994A&A...290..553R} as implemented in the codes of \cite{2001ApJ...557..389U} and \cite{1995A&A...299..563H} for the optically thick regime. In the solar case, at the present time we have rather little observational motivation for attempting to model the UV/optical lines and continuum, with the exception of a few, due to the paucity of observational data. But in the stellar case this is obviously of great value - not only the overall shape of the spectrum, but the detailed profiles of lines can be used to diagnose the flaring plasma. In the solar case, a small number of lines have been observed and modeled in some detail. These include H$\alpha$ line, the profile and time evolution of which has been studied by many authors, e.g. \cite{2009A&A...499..923K}. Careful study of the line profiles offers the possibility of discriminatory tests for the nature of the chromospheric energy input, including the spectral index of the electron spectrum that excites the H$\alpha$ radiation and its fine structuring in time. Other lines which have been modeled in some detail include Ca II K \citep{1993A&A...274..917F}  and the Ca II $\lambda$8542 line \citep{2003JKAS...36S..49D,2008A&A...490..315B}, i.e. those which have received observational attention.

Semi-empirical models of flaring atmospheres are much used, and certain of them have become standards \citep[e.g. the `F1' model of][]{1980ApJ...242..336M}. They are certainly useful, but are constructed by analyzing data from different times of different flares, so can only give some kind of averaged view of a flare chromosphere. 
They tend not to represent the impulsive phase very well, both because the input spectroscopic information is so limited and because the physics may be more complicated.
Such models are in the single-fluid approximation with $T_e = T_p$, and therefore do not handle interesting parts of the plasma physics.

\section{More exotic flaring objects}
The wide variety of astrophysical sources offers the possibility to study magnetically-driven energy storage and release in many situations, not just isolated stars. These include the magnetised star plus disk configuration of a young stellar object  \citep{2005ApJS..160..469F}, and the binary configuration of RS CVns, in which the two components are interacting and possibly magnetically linked \citep{1985IAUS..107..281U}. The variety of configurations and activity levels may allow us to investigate questions like what determines the energy and evolution of a flare in a given magnetic configuration? Is it just the overall size or mean magnetic field, or are there other restrictions placed by the accessible topologies of the system? And what determines whether a flare results in heating primarily, or in particle acceleration? Is it only a function of the total energy, or some other parameter such as the mean density and temperature (i.e. collisionality) of the plasma in which the primary energy release takes place? Is the Sun unusual in converting stored magnetic energy efficiently into the KE of non-thermal electrons?

The pre-main sequence stars, which were the main target of the COUP survey mentioned above, represent ``normal'' stars in probably their most magnetically-active phase. Their X-ray luminosity can be $>10^4$ times that of a solar flare's X-ray luminosity \citep[Figure 8 in][]{2008ApJ...688..418G} which is in turn about 10\% of the total solar flare luminosity.  As discussed by Getman in this session, flares from both disk bearing and diskless COUP stars tend to last 10-100 times longer than solar events, and also involve loops which are 100-1000 times as long as in the solar case - many times the radius of the star (length being determined from a cooling timescale analysis). One particular PMS object, DQ Tau, offers an unusual opportunity to study repeated flaring as it is in a binary with a period of just under 16 days and undergoes mm band flaring at periastron \citep{2010A&A...521A..32S}, attributed to interactions of the magnetospheres of the binary components - the typical loop lengths derived from the above-mentioned analyses are comparable with the periastron separation, mm and X-ray flares are causally related, and the energy released is comparable with what should be expected from reconnection in colliding magnetospheres (Getman et al. 2010 submitted).

More energetic yet are the magnetars - highly magnetised pulsars - which also show extreme flare outputs but represent physical conditions which differ so much from normal MS stars that it is not clear what can be learned from their study which is of direct relevance to interpreting the radiation signature of solar or stellar flares.

\section{Proposals for the future}
At the end of this splinter session, a set of questions was proposed for future study, and some general comments made. 
\begin{itemize}
\item When is a feature seen in a light curve to be classified as a flare?
\item Do all major flares (solar and stellar) show an impulsive phase?
\item What is the energy conversion process during the impulsive phase?
\item What do the timescales in the energy-release phase represent?
\item What is the relationship between microflaring and coronal heating?
\item Are scaled-up versions of the solar mechanisms always appropriate for attempting to understand stellar flares?
\item Can ideas of topology be usefully applied to stellar as well as solar flares?
\item What is the role of accelerated particles in stellar flare energy budgets?
\end{itemize}

We still have a lot to learn about the theory of solar flares, so stellar flare physicists are cautioned to investigate models beyond the `standard flare model'. This model, embodied in the `Carmichael-Sturrock-Hirayama-Kopp-Pneuman' cartoon, is one of many many scenarios possible for flares (see e.g.
 \url{http://solarmuri.ssl.berkeley.edu/~hhudson/cartoons/}). The CSHKP cartoon may do a decent job at encapsulating the gradual phase of a flare but was never intended to describe the impulsive energy release. We do not know what triggers the solar flare instability, nor for certain how energy is transported through the flaring atmosphere (the long-standing `number problem' and `return current problem' with the electron beam model \citep[e.g.][]{1977SoPh...52..117B,1976SoPh...48..197H,2008ApJ...675.1645F} are not solved satisfactorily). Proper kinetic modeling of the plasma processes, though crucial for some areas (e.g. coherent radio emission) is not nearly as widespread as it should be, and the treatment of chromospheric processes assumes that the plasma is a single Maxwellian fluid, despite the fact that collisionally-maintained LTE is unlikely in  the extreme energy inputs received by small areas of the chromosphere. 
 
Observationally, to understand better the solar processes, we advocate an emphasis on optical spectroscopy of the chromosphere. Relatively easily observed compared to hard X-rays, white-light (and UV-continuum) observations provide a view of the total energy of a flare complementary to that provided by HXR radiation.
There is little effort to conduct studies of chromospheric flares from space, though there has been notable success in targeted programs using TRACE, and serendipitous observations with \textit{Hinode}. It is to be expected that the \textit{Solar Dynamics Observatory} will also reveal the detailed morphological evolution of ribbons, and possibly give some insight also into the structure of the bright ribbon sources. At its most basic, continuum spectroscopic information is required for a proper assessment of the total radiative intensity in flares (which has so far been obtained only in a small number of events), and  to determine if there are white-light properties during solar flares not predicted by RHD models, as is the case for the hot, blackbody component of M dwarf flares. Of particular interest is the identification \citep{1989SoPh..121..261N} of the Paschen jump (indicating the presence of the Paschen continuum from free-bound emission in the optical part of the spectrum) and flare-related changes to the Balmer and Paschen jumps in the flare spectrum, indicating variations in ionization fraction. Enhanced continuum shortwards of the Paschen jump may also suggest increased H$^-$ opacity, and photospheric excitation. This basic measurement has been made in very few flares \citep[see e.g.][]{1983SoPh...85..285N,1990ApJ...360..715M}, and usually not for the strongest impulsive-phase optical flare kernels, but has major implications for the energy transport model \citep{1989SoPh..121..261N,2008ApJ...675.1645F}.

\acknowledgements{LF would like to thank the CS16 Program Committee for selecting this Splinter Session to run, and all the participants for a lively and stimulating discussion. She would also like to acknowledge financial support from the European Community \& Research Training Network project SOLAIRE(MTRN-CT-2006-035484), from UK STFC Rolling Grant ST/F002637/1, and from the International Space Sciences Institute, Bern whose support of the `Chromospheric Flares' team led directly to this Splinter Session.}

\bibliographystyle{asp2010}
\bibliography{fletcher_l}

\end{document}